\documentclass[12pt,epsf]{article}

\def\be{\begin{equation}}
\def\ee{\end{equation}}
\def\bea{\begin{eqnarray}}
\def\eea{\end{eqnarray}}

\def\be{\begin{equation}}
\def\ee{\end{equation}}
\def\bea{\begin{eqnarray}}
\def\eea{\end{eqnarray}}
\def\half{\frac{1}{2}}

\def\case#1/#2{\textstyle\frac{#1}{#2}}
\def\k0{\kappa_{0}}

\begin{document}
\begin{titlepage}

\vspace{.7in}

\begin{center}
\Large
{\bf Exact Inflationary Solutions from a Superpotential}\\
\vspace{.7in}
\normalsize
\large{ 
Alexander Feinstein
}\\
\normalsize
\vspace{.4in}

{\em Dpto. de F\'{\i}sica Te\'orica, Universidad del Pa\'{\i}s Vasco, \\
Apdo. 644, E-48080, Bilbao, Spain}\\
\vspace{.2in}
\end{center}
\vspace{.3in}
\baselineskip=24pt

\begin{abstract}
\noindent We propose a novel, potentially useful   generating technique
for constructing exact  solutions of inflationary scalar field cosmologies with non-trivial
potentials. The generating scheme uses the so-called superpotential and is inspired by
 recent studies of similar equations in supergravity. Some exact solutions are derived, and
the physical meaning of the superpotential in these models is clarified. 
\end{abstract}

\vspace{.3in}

\end{titlepage}

Inflaton driven cosmological expansion,
although not without some problems, still represents probably our best bet to describe the 
evolution of the early universe. 
On the other hand, the dynamics of the present day universe, we used to believe until now, is
 correctly described by a standard decelerating model.
The new observational evidence \cite{perl,gar}, however,   points in favour of spatially flat weakly accelerating universe 
even at present. This weak acceleration must be induced by a  matter with
 negative pressure.

Some new models \cite{zlat, albr} were  proposed recently to account for the  
presence and for a possible present time dominance of energy density with negative pressure.
Apart from  usual radiation or matter, these models contain an additional ingredient
either in the form of a constant vacuum
stress, or quintessence, usually taken to be a  scalar field slowly rolling
down some non-trivial potential. In these model universes both the ordinary matter, the
adiabatic perfect fluid, and the scalar field, are equally important, and one of the main
enigmas to be resolved in this context is why the energy density of the additional quintessence matter almost
co-incides with the usual matter density today \cite{zlat,barr}. 
It follows, that the main player in these models becomes the shape of the scalar field
potential, which due to scaling properties of the energy density of the scalar field
\cite{rat, ferr, lid} determines much of the model phenomenology. In these scenarios one may
divide the history of the universe into distinct periods depending on which source,
the scalar field, or matter (radiation)
dominates the expansion. Initially, to allow for the fast inflation, the universe should be
dominated by the scalar  field, then the radiation and matter energies take over, and then 
at present, the universe must start accelerating, signalling that the scalar field energy  becomes 
important once again. 

During the periods of  the scalar field domination, one may discard the other
forms of matter and just concentrate on the behaviour of the scalar field driven expansion,
therefore, the study of the scalar field  models with different self-interaction terms
 becomes of great interest by its own right.
Although many qualitative and numerical studies
of scalar field models were undertaken during the last years
due to the advent of inflationary cosmology, the number of exact
solutions of Einstein field equations with self-interacting scalar fields, even in the presence of
high symmetry, is rather poor.

The main purpose of this Letter is to present a {\em simple} generating technique  for solving
coupled Einstein-scalar field equations with non-trivial potentials. We  limit our
discussion  to the  case of spatially flat isotropic universe, anticipating, that
in more complex situations, either in presence of spatial curvature, or with extra
matter fields, the method we employ doesn't seem to be too promising.

To this end  we consider  spatially flat FRW line element given by:

\begin{equation}
ds^2  = dt^2  - \exp (2A)(dx^2  + dy^2  + dz^2 ),
\end{equation}
where $A$ is a function of time alone.

The Einstein Equations with the scalar field stress tensor  are given by:

\be
R_{\mu\nu}=\phi_{\mu}\phi_{\nu}- g_{\mu\nu}V(\phi),
\label{ee}
\ee
and may be cast after some algebraic manipulations into  the two 
following {\em independent}  equations
\bea
2\ddot A + \dot \phi ^2  &=& 0 \label{eq1}\\
3\dot A^2  - \frac{1}{2}\dot \phi ^2  &=&   V \label{eq2}
\eea
 The energy conservation equation is identically satisfied when (\ref{eq1}) 
and (\ref{eq2}) hold.
We have set the constants appearing in the Einstein equations to $1$,
and are  using the usual GR normalisation
for the scalar field  as follows from (2).

Given a potential $V(\phi)$ one may try to solve these coupled differential equations.
 This happens to be  rather difficult a task.
 Some years ago Ellis and Madsen \cite{ellis}
have proposed  a way of ``solving" these equations by first assuming
some particular form for the function $A(t)$. Differentiating $A(t)$ 
twice one may readily read $\dot\phi$ from  (\ref{eq1}), and then  you obtain $V(t)$ from 
 (\ref{eq2}). In other words, you just assume the expansion you wish,
 and then you  read the potential after doing some  calculus, ``solving" the
Einstein equations from  ``left to right".
 Since what one is really  interested in is the potential as a function of $\phi$ you must invert
$\phi(t)$ into $t(\phi)$, and then substitute this into $V(t)$ to finally get $V(\phi(t))$.
This may be easily done for some simple choices of the scale factor 
$a(t) \equiv \exp{A(t)}$, those like
 $a(t)\propto  t^{n}$,
or $a(t)\propto \exp{(kt)}$, however if the scale factor is more involved the procedure 
 of  inversion of $t(\phi)$ may become very complicated.
The main flaw in this scheme,  however, is that even if one manages to invert 
the above expressions, the potential $V(\phi)$ one finally finds may happen to be completely irrelevant.

A different approach to solve this system of equations is suggested by their similarity 
to those ones arising in supergravity \cite{dewo}. Here, one
introduces the so-called superpotential $W(\phi)$, such that 
\be
V=-2(\frac{\partial{W}}{\partial\phi})^{2}+3W^{2}, \label{super}
\ee
along with
\be
W(\phi)=\dot{A} \label{A}
\ee
and
\be
\dot\phi=-2\frac{\partial{W}}{\partial{\phi}} \label{phi}
\ee

In 5-D gauged supergravity, the potentials of the form (\ref{super}) occure
naturally, but here, we will merely use $W(\phi)$ as a solution generation function.
It is straightforward to see that the Einstein equations are immediately satisfied by
this choice, and moreover, the solution spaces of both sets of equations
(\ref{eq1}, \ref{eq2}) on one hand, and  (\ref{super}, \ref{A} and \ref{phi}), on the
other, are identical.
The strategy for solving the equations (\ref{eq1} and \ref{eq2}) then might be the following:
Choose the superpotential $W(\phi)$. 
This immediately defines the shape of the potential, being 
the principle advantage of this generating scheme.  Next, from (\ref{phi}) we can find
$t(\phi)$ and  invert it to $\phi(t)$. The function $A(t)$, on the other hand, is then obtained by the
quadrature from (\ref{A}). The inversion is still a technical nuisance of the method, nevertheless, 
the transparency of the form of the potential is worth the effort.

We now apply the above outlined  scheme to obtain some exact solutions.  First,
 we check that the superpotential
$W=c\exp{k\phi}$ which defines the  exponential potential
$$ 
V(\phi)=c^{2}(3-2k^{2})\exp({2k\phi})
$$
gives an expected  power low scale factor \cite{exp}
$$
a=t^{\frac{1}{2k^{2}}},
$$
with the scalar field being
$$
\phi=\frac{1}{k}[\log{t}+\log{2ck{^2}}]
$$

Interestingly enough, a simple constant shift in $W(\phi)$
$$
W=c\exp{k\phi}+b, 
$$
changes  the  potential to:
$$
V(\phi)=c^{2}(3-2k^{2})\exp({2k\phi})+6cb\exp{(k\phi)}+3b^{2},
$$
with the scalar field being unchanged, but introducing a new behaviour into the
scale factor
$$
a=\exp{(bt)}t^{\frac{1}{2k^{2}}}
$$
It is as if we have added a cosmological constant to the potential and the scale factor,
which had a power low behaviour before this addition, has been enhanced ( multiplied) by an
exponential de Sitter term.
 
We now consider  a power low behaviour for $W(\phi)$:
$$ 
W(\phi)=k\phi^{n}+b 
$$
The scalar field potential is given then by:
\be
V(\phi)=3k^{2}\phi^{2n}-2(nk)^{2}\phi^{2n-2} +6kb\phi^{n}+3b^{2} \label{v1},
\ee
and integrating ( and inverting) (\ref{phi}) we get for the scalar field ( for $n\neq{2}$)
\be
\phi=[2kn(n-2)t]^{\frac{1}{2-n}},
\ee
 Resolving (\ref{A}) for the metric function we find
\be
A=k(1-\half{n})[2kn(n-2)]^{\frac{n}{2-n}}t^{\frac{2}{2-n}} +bt
\ee
In the case $n=2$ we have
\be
\phi=\exp{(-4kt)}
\ee
and 
\be
A=-\half\exp{(-8kt)}+bt
\ee

Needless to say that the  potential given by (\ref{v1}) is quite rich and represents
many interesting particular cases: for $n=1$ and $b=0$ we have a typical case of quadratic
potential, whereas for $n=2$ and $b=0$ we have a quartic potential resulting in a double exponential
scale factor $a$. Tracker type solutions may be  obtained  for negative  powers of $n < 0$
in $W(\phi)$ \cite{zlat}. In terms of the 
superpotential, the scaling index $\xi$  of Ref. \cite{ferr} is given by
\be
\xi=1-\frac{2}{3}\left[\frac{W'(\phi)}{W(\phi)}\right]^{2},
\ee
here prime denotes derivative with respect to $\phi$.
It is easy to see that the scaling $\xi$ is constant only for a constant or for an exponential
superpotential which results in exponential or power low behaviour of the scale factor
respectively.
Alternatively, if we write the scaling behaviour as $\rho_{\phi}\propto 1/a^{m}$, we then  have
that $$
m=4 \left[\frac{W'(\phi)}{W(\phi)}\right]^{2},
$$
here one must assume that the  ``adiabatic" index of the perfect fluid model describing the scalar
field is constant \cite{ferr}. The identification of the perfect fluid and the scalar field,
 just to recall,
is achieved by identifying the velocity potential of the fluid with the scalar field.

The scalar field energy density can also be expressed in terms of the superpotential and simply
becomes
\be 
\rho=\half \dot \phi ^2+V(\phi)=3W^{2}(\phi),
\ee
hence, we arrive at quite an amusing result that 
 the superpotential $W(\phi)$ has a rather simple physical meaning as the 
{\em square root of the energy density of the scalar field}. 

Phenomenologically, the cosmological models accelerate or decelerate depending on the 
sign of the second derivative of the scale factor. Expressing the acceleration conditions in terms of 
$W$ gives accelerating models whenever
$$
W^{2}-2 \left[\frac{W'(\phi)}{W(\phi)}\right]^{2}>0,
$$
 otherwise we have a ``standard" decelerating expansion, and it is not difficult to see
what happens with each model. We have checked that depending on the power $n$ and on the constants 
$k$ and $b$ one may construct with  the potential (8)  models to ones taste: ever inflating models,
ever decelerating models, or models where the deceleration and inflation interchange several times.

Our final example, just before closing, is the potential obtained from $W=k\cosh{b\phi}$,
\be
V=3k^{2}\cosh^{2}{(b\phi)}-2(kb)^{2}\sinh^{2}{(b\phi)}
\ee
In this case the scalar field is 
\be
\phi=\frac{1}{b}Arctanh{[\exp{(b^{2}kt)}]},
\ee
and the metric function $A$ may be integrated to give 
\be
A= \half k \left(\coth ^{2}(b^{2}t)+1\right)t,
\ee
and  of course one may proceed  further with generating variety of models.
Therefore, it becomes relatively easy to construct exact  solutions for
different shapes of potentials. Moreover, since in principle one may express all
the relevant dynamical information about scalar field models via the function $W(\phi)$
and its first derivative,
this function may serve as an interesting dynamical variable to study the models analytically
as well as doing some numerics, but this is away from the scope of this note.

To sum up, we have presented simple generating procedure to construct exact
solutions for scalar field isotropic and spatially flat
cosmologies with potentials of different shapes. It is hoped that
this method will become fruitful for generating and studying models of physical interest,
as well as it is hoped that the  superpotential function $W(\phi)$ might turn a useful tool 
for studying dynamical properties of these models.

This research is
supported by the University of the Basque Country Grant
UPV 122.310-EB150/98, General University Research Grant UPV172. 310-G02/99
 and Spanish Science Ministry Grant PB96-0250.

\end{document}